\documentclass{natureprintstyle}
\usepackage{ifthen}
\usepackage{graphicx}
\usepackage{amsmath}
\usepackage{amssymb}

\newboolean{SubmittedVersion}
\setboolean{SubmittedVersion}{False}
\DeclareMathAlphabet{\mathpzc}{OT1}{pzc}{m}{it}

\def\Rb87{^{87}\text{Rb}}
\def\Na23{^{23}\text{Na}}
\def\Li6{^{6}\text{Li}}

{\catcode`\|=\active
  \gdef\Braket#1{\left<\mathcode`\|"8000\let|\BraVert {#1}\right>}}
\def\BraVert{\egroup\,\mid@vertical\,\bgroup}

\begin{document}

\title{Locating influential nodes via dynamics-sensitive centrality}
\author{Jian-Hong Lin$^1$, Qiang Guo$^1$, Jian-Guo Liu$^1$* Tao Zhou$^{2,3}$*}
\maketitle

\begin{affiliations}
$^1$Research Center of Complex Systems Science, University of Shanghai for Science and Technology, Shanghai 200093, PR China;
$^2$CompleX Lab, Web Sciences Center, University of Electronic Science and Technology of China, Chengdu 611731, PR China;
$^3$Big Data Research Center, University of Electronic Science and Technology of China, Chengdu 611731, PR China.
\end{affiliations}

\begin{abstract}
With great theoretical and practical significance, locating influential nodes of complex networks is a promising issues. In this paper, we propose a dynamics-sensitive (DS) centrality that integrates topological features and dynamical properties. The DS centrality can be directly applied in locating influential spreaders. According to the empirical results on four real networks for both susceptible-infected-recovered (SIR) and susceptible-infected (SI) spreading models, the DS centrality is much more accurate than degree, $k$-shell index and eigenvector centrality.
\end{abstract}

~\\

Spreading dynamics represents many important processes in nature and society~\cite{Zhou2006,Pastor2014}, such as the propagation of computer virus~\cite{Kephart1997} and traffic congestion~\cite{Li2015}, the reaction diffusion~\cite{Colizza2007}, the spreading of infectious diseases~\cite{Keeling2008} and the cascading failures~\cite{Motter2004}. The estimation of nodes' spreading influences can help in hindering the epidemics or accelerating the innovation~\cite{Pei201302}, and similar methods can be further applied in identifying the influential spreader in social networks~\cite{Gonzalez-Bail¨®n2011}, quantifying the influence of scientists and their publications~\cite{Zhou201201}, evaluating the impacts of injection points in the diffusion of microfinance~\cite{Banerjee2013}, finding drug targets in directed pathway networks~\cite{Csermely2013}, predicting essential proteins in protein interaction networks~\cite{Li2012}, and so on.

The significance of this issue triggers a variety of novel approaches in identifying influential spreaders in networks, which can be roughly categorized into three classes. Firstly, some scientists argued that the location of a node is more important than its immediate neighbors, and thus proposed $k$-shell index~\cite{Kitsak201001,Castellano2012} and its variants~\cite{Zeng2013,Liu201310,Lin2014} as indicators of spreading influences. Secondly, some scientists quantified a node's influence only accounting for its local surroundings~\cite{Chen2012,Chen201302,Pei201401}. Thirdly, some scientists evaluated nodes' influences according to the steady states of some introduced dynamical processes, such as random walk~\cite{L201101,Li2014} and iterative refinement~\cite{Ren201401}.

The above-mentioned approaches only take into account the topological features, while recent experiments indicate that the performance of structural indices is very sensitive to the specific dynamics on networks~\cite{Borge-Holthoefer201201,Borge-Holthoefer201202,Liu2014}. For example, when the spreading rate is very small, the degree usually performs better than the eigenvector centrality~\cite{Borgatti2005} and $k$-shell index~\cite{Kitsak201001}, while when the infectivity is very high, the eigenvector centrality is the best one among the three (see figures 1 and 2, with details shown later). To the best of our knowledge, there are few works taking into account the properties of the underlying spreading dynamics~\cite{Klemm201202,Li201202,Bauer2012}. Via a Markov chain analysis, Klemm \emph{et al.}~\cite{Klemm201202} suggested that the eigenvector centrality can be used in estimating nodes' dynamical influences in the susceptible-infected-recovered (SIR) spreading model (also called susceptible-infected-removed model)~\cite{Hethcote2000}. Li \emph{et al.}~\cite{Li201202} provided complementary explanation of the suitability of eigenvector centrality based on perturbation around the equilibrium of the epidemic dynamics and discussed the limitations of eigenvector centrality for homogeneous community networks. Both the above two works did not pay enough attention to the specific parameters in the spreading models, and thus their suggested index only works well in a limited range of the parameter space. Bauer and Lizier~\cite{Bauer2012} directly counted the number of possible infection walks with different lengths. Their method is very effective but less efficient due to the considerable computational cost, in addition, for the fundamental complexity in counting the number of paths connecting two nodes, their method can not be formulated in a compact analytical form.

In this paper, we describe the infectious probabilities of nodes by a matrix differential function that accounts both topological features and dynamical properties. Accordingly, we propose a dynamics-sensitive (DS) centrality, which can be directly applied in quantifying the spreading influences of nodes. According to the empirical results on four real networks, for both the SIR model~\cite{Hethcote2000} and the susceptible-infected (SI) model~\cite{Barthelemy2004,Zhou2006b}, the DS centrality is more accurate than degree, $k$-shell index and eigenvector centrality in locating influential nodes. The method proposed in this paper can be extended to other Markov processes on networks.

\section{Dynamics-Sensitive Centrality}

~\\

An undirected network $G=(V,E)$ with $n=|V|$ nodes and $e=|E|$ edges could be described by an adjacent matrix $\mathbf{A}=\left\{a_{ij}\right\}$ where $a_{ij}=1$ if node $i$ is connected with node $j$, and $a_{ij}=0$ otherwise. $\mathbf{A}$ is binary and symmetric with zeros along the main diagonal, and thus its eigenvalues are real and can be arrayed in a descending order as $\lambda_{1}{\geq}\lambda_{2}{\geq}{\ldots}{\geq}{\lambda}_{n}$. Since $\textbf{A}$ is a symmetric and real-valued matrix, it can be factorized as $\mathbf{A}=\mathbf{Q}\mathbf{\Lambda}\mathbf{Q}^{T}$, where $\mathbf{\Lambda}=\rm{diag}(\lambda_{1},\lambda_{2},{\ldots},{\lambda}_{n})$, $\textbf{Q}=[\textbf{q}_{1},\textbf{q}_{2},\ldots,\textbf{q}_{n}]$ and $\textbf{q}_{i}$ is the eigenvector corresponding to $\lambda_{i}$.

Considering a spreading model where an infected node would infect its neighbors with spreading rate $\beta$ and recover with recovering rate $\mu$ (see \textbf{Materials and Methods} for details). We denote $\textbf{x}(t)$ ($t\geq 0$) as an $n\times1$ vector whose components are the probabilities of nodes to be ever infected no later than the time step $t$, and then $\textbf{x}(t)-\textbf{x}(t-1)$ ($t>1$) is the probabilities of nodes to be infected at time step $t$. Notice that, $\textbf{x}(t)$ is the cumulative probability that can be larger than 1, and we use the term \emph{probability} just for simplicity. For example, if $i$ is the only initially infected node, then $x_{i}(0)=1$ and $x_{j\neq{i}}(0)=0$. In the first time step, $\textbf{x}(1)=\beta\mathbf{A}\textbf{x}(0)$, and for $t>1$, we have (see the derivation in \textbf{Materials and Methods})
\begin{equation}\label{equation1}
\textbf{x}(t)-\textbf{x}(t-1)={\beta}\mathbf{A}[{\beta}\mathbf{A}+(1-\mu)\mathbf{I}]^{t-1}{\textbf{x}(0)},
\end{equation}
where $\textbf{I}$ is an $n\times{n}$ unit matrix. Denoting $\mathbf{H}={\beta}\mathbf{A}+(1-\mu)\mathbf{I}$, then ${\beta}\mathbf{A}\mathbf{H}^{t-1}{\textbf{x}(0)}$ represents the probabilities of nodes to be infected at time step $t$, and thus the probabilities of nodes to be ever infected no later than $t$ can be rewritten as
\begin{equation}\label{equation2}
\textbf{x}(t)=\sum_{r=2}^{t}[\textbf{x}(r)-\textbf{x}(r-1)]+\textbf{x}(1)=\sum_{r=0}^{t-1}{\beta\mathbf{A}\mathbf{H}^{r}\textbf{x}(0)}.
\end{equation}

We define $S_i(t)$ the spreading influence of node $i$ at time step $t$, which can be quantified by the sum of infected probabilities of all nodes, given $i$ the initially infected seed. According to Eq. (2), the infected probabilities can be written as
\begin{equation}
\textbf{x}(t)=\sum_{r=0}^{t-1}{\beta\mathbf{A}\mathbf{H}^{r}\textbf{e}_i},
\end{equation}
where $\textbf{e}_i=(0,\cdots,0,1,0,\cdots,0)^T$ is an $n\times 1$ vector with only the $i$th element being 1. As all elements other than the $i$th one of $\textbf{e}_i$ are zero, $\textbf{x}(t)$ is indeed the sum of all the $i$th columns of $\beta\mathbf{A}, \beta\mathbf{A}\mathbf{H}, \cdots, \beta\mathbf{A}\mathbf{H}^{t-1}$. Given $\textbf{x}(0)=\textbf{e}_i$, $S_i(t)$ is defined as the sum of all elements of $\textbf{x}(t)$, which is equal to the sum of all elements in the $i$th columns of $\beta\mathbf{A}, \beta\mathbf{A}\mathbf{H}, \cdots, \beta\mathbf{A}\mathbf{H}^{t-1}$, as
\begin{equation}
S_i(t)=\left[ \left( \beta\mathbf{A} + \beta\mathbf{A}\mathbf{H} + \cdots + \beta\mathbf{A}\mathbf{H}^{t-1} \right)^T \textbf{L} \right]_i,
\end{equation}
where $\textbf{L}=(1,1,\cdots,1)^T$ is an $n\times1$ vector whose components are all 1. Obviously, $\mathbf{A}^T=\mathbf{A}$, $\mathbf{H}^T=\mathbf{H}$ and $\mathbf{A}\mathbf{H}=\mathbf{H}\mathbf{A}$, then the spreading influence of all nodes can be described by the vector
\begin{equation}\label{equation3}
\textbf{S}(t)=\sum_{r=0}^{t-1}{\beta\mathbf{A}\mathbf{H}^{r}\textbf{L}}.
\end{equation}
Notice that, $\sum_{r=0}^{t-1}{\beta\mathbf{A}\mathbf{H}^{r}\textbf{L}}=\sum_{r=0}^{t-1} \beta\mathbf{A}\mathbf{H}^{r}\left( \sum^n_{i=1}\textbf{e}_i \right) = \sum^n_{i=1} \sum_{r=0}^{t-1} \beta\mathbf{A}\mathbf{H}^{r} \textbf{e}_i$, and $\sum_{r=0}^{t-1} \beta\mathbf{A}\mathbf{H}^{r} \textbf{e}_i$ is the infected probabilities of all nodes given node $i$ the only initially infected seed according to Eq. (2), so $\textbf{S}(t)$ can also be roughly explained as the sum of infected probabilities over the $n$ cases with every node being the infected seed once. This relationship shows an underlying symmetry, that is, in an undirected network, the node having higher influence is also the one apt to be infected. The readers are warned that such conclusion is not mathematically rigorous since we have ignore the complicated entanglement by allowing the elements of $\textbf{x}(t)$ being larger than 1.

According to the Perro-Frobenius Theorem \cite{Hom1985}, the eigenvectors of $\mathbf{H}$ is the same to the ones of $\mathbf{A}$ and $\beta\lambda_{i}+1-\mu$ is the $i$th eigenvalue of $\mathbf{H}$, corresponding to $\textbf{q}_{i}$. When $\beta\lambda_{1}+1-\mu<1$, i.e. $\beta/\mu<1/\lambda_{1}$ (for the case $\mu\neq0$), $\mathbf{H}^{t}\textbf{L}$ could converge to null vector when $t\rightarrow \infty$ and $\mathbf{S}(t)$ could be written by the following way
\begin{equation}\label{equation4}
\mathbf{S}(t)=\beta\mathbf{A}(\mathbf{I}-\mathbf{H})^{-1}\textbf{L}=[(\beta/\mu)\mathbf{A}+(\beta/\mu)^2\mathbf{A}^2+\cdots+(\beta/\mu)^t\mathbf{A}^t]\textbf{L}.
\end{equation}
For SIR model, without loss of generality, we set $\mu=1$, and then
\begin{equation}
\mathbf{S}(t)=(\beta\mathbf{A}+\beta^2\mathbf{A}^2+\cdots+\beta^t\mathbf{A}^t)\textbf{L},
\end{equation}
where $(\mathbf{A}^{t}\textbf{L})_{i}$ counts the total number of walks of length $t$ from node $i$ to all nodes in the network, weighted by $\beta^{t}$ that decays as the increase of the length $t$. As $\mathbf{S}(t)$ quantifies nodes' spreading influences, we call it dynamics-sensitive (DS) centrality, where the term \emph{dynamics-sensitive} emphasizes the fact that $\mathbf{S}(t)$ is determined not only by the network structure (i.e., $\mathbf{A}$), but also the dynamical parameters (i.e., $\beta$ and $t$). In particular, when $t=1$, the initially infected node only has the chance to infect its neighbors and ${S}_{i}(1)=(\beta\mathbf{AL})_{i}$ with $(\mathbf{AL})_{i}$ being exactly the degree of node $i$. When $\mu=0$ (corresponding to the SI model) or $\beta\geq1/\lambda_{1}$, $\mathbf{S}(t)$ would be infinite when $t\rightarrow\infty$, which could not reflect the spreading influences. In fact, as we allow the infected probability of a node to be cumulated and exceed 1, the DS centrality may considerably deviate from the real spreading influences at large $\beta$ and large $t$. It is because that our theoretical deviation contains an underlying approximation that $1-(1-\beta)^m\approx m\beta$, where the left-hand side characterizes the real spreading process with any infected probability smaller than 1 and the right-hand side could exceed 1 when $\beta$ and $m$ are large. Since $m$ denotes the number of contacts from infected nodes, the larger $t$ will result in larger $m$ before the end of epidemic spreading. Notice that, our main goal is to find out the ranking of spreading influences of nodes, namely to identify influential nodes, and thus we are still not aware of the impacts on the ranking from the above deviation because every node's influence is over estimated. Fortunately, as we will show later, the DS centrality performs much better than other well-known indices for a very broad ranges of $\beta$ and $t$ that cover most practical scenarios.

\begin{figure*}[ht]
\center\scalebox{0.55}[0.55]{\includegraphics{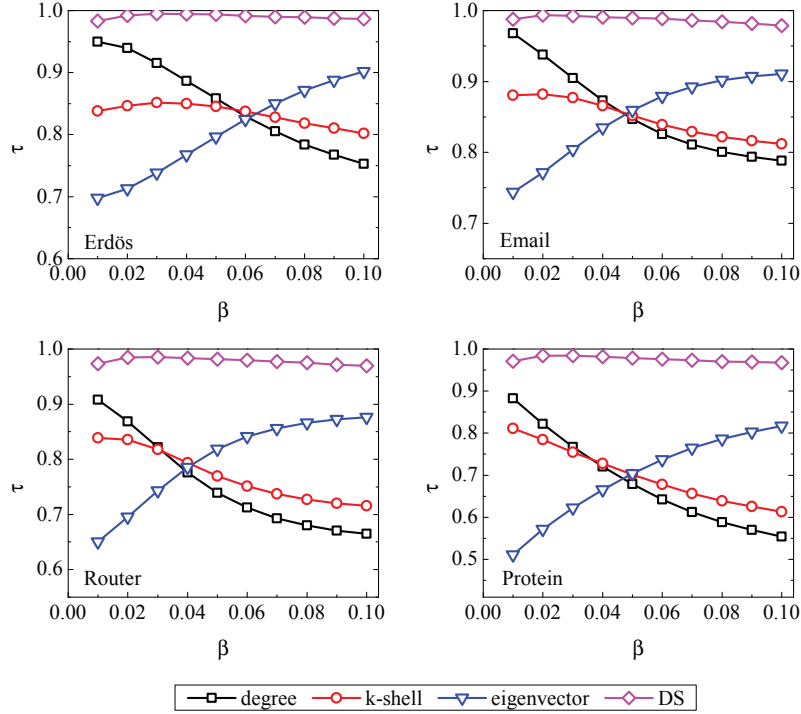}}
\caption{The accuracy of four centrality measures in evaluating nodes' spreading influences according to the SIR model ($\mu=1$) in the four real networks, quantified by the Kendall's Tau. The spreading rate $\beta$ varies from 0.01 to 0.10, and the time step is set as $t=5$. Each data point is obtained by averaging over $10^4$ independent runs.}
\end{figure*}

\section{Results}

~\\

We test the performance of DS centrality in evaluating the nodes' spreading influences according to the SIR model and SI model, with varying spreading rate $\beta$. Four real networks, including a scientific collaboration network, an email communication network, the Internet at the router level and a protein-protein interaction network, are used for the empirical analysis (see data description in \textbf{Materials and Methods}), and three well-known indices, including degree, $k$-shell index and eigenvector centrality, are used as benchmark methods for comparison (see \textbf{Materials and Methods} for the definitions of those indices). Given the time step $t$, the spreading influence of an arbitrary node $i$ is quantified by the number of infected nodes (for SI model) or the number of infected and recovered nodes (for SIR model) at $t$, where the spreading process starts with only node $i$ being initially infected. Here we use Kendall's Tau $\tau$~\cite{Kendall1938} to measure the correlation between nodes' spreading influences and the considered centrality measure, where $\tau$ is in the range $[-1,1]$ and the larger $\tau$ corresponds to the better performance (see \textbf{Materials and Methods} for the definition of $\tau$).

As shown in Fig. 1, the Kendall's Tau $\tau$ for the DS centrality is between 0.968 and 0.995 for $\beta \in [0.01,0.1]$, indicating that the ranking lists generated by the DS centrality and the real SIR spreading process are highly identical to each other. In comparison, the DS centrality performs much better than degree, $k$-shell index and eigenvector centrality. As shown in Fig. 2, similar result is also observed for the SI model where the DS centrality performs much better than others. The results for larger $\beta$ and $t$ are respectively shown in Fig. S1 and Fig. S2 of \textbf{Supplementary Information}, where the DS centrality still performs the best.

Since $\mathbf{A}$ is a symmetric, real-valued matrix, the DS centrality $\textbf{S}(t)$ can be written in the following way by decomposing $\mathbf{A}$
\begin{equation}
S_{i}(t)=m_{1}q_{1i}\sum_{j=1}^{n}q_{1j}+\sum_{r=2}^{n}m_{r}q_{ri}\sum_{j=1}^{n}q_{rj},
\end{equation}
where $m_{r}=\beta\lambda_{r}[1-(\beta\lambda_{r}+1-\mu)^t](\mu-\beta\lambda_{r})^{-1}$ for $1\leq{r}\leq{n}$. Rewriting Eq. (8) into
\begin{equation}
\frac{S_{i}(t)}{m_1}=q_{1i}\sum_{j=1}^{n}q_{1j}+\sum_{r=2}^{n}\frac{m_{r}}{m_{1}}q_{ri}\sum_{j=1}^{n}q_{rj}.
\end{equation}
With the increase of $t$ and $\beta$, $\frac{m_{r}}{m_{1}}$ will converge to $0$, and thus the ranking lists generated by $\textbf{S}(t)$ will be identical to $\textbf{q}_{1}$, which is exactly the same to the eigenvector centrality. This relationship is in accordance with the results presented in Fig. S1 and Fig. S2, where the difference between the eigenvector centrality and DS centrality gets smaller as the increase of $\beta$ and $t$.

\begin{figure*}[ht]
\center\scalebox{0.55}[0.55]{\includegraphics{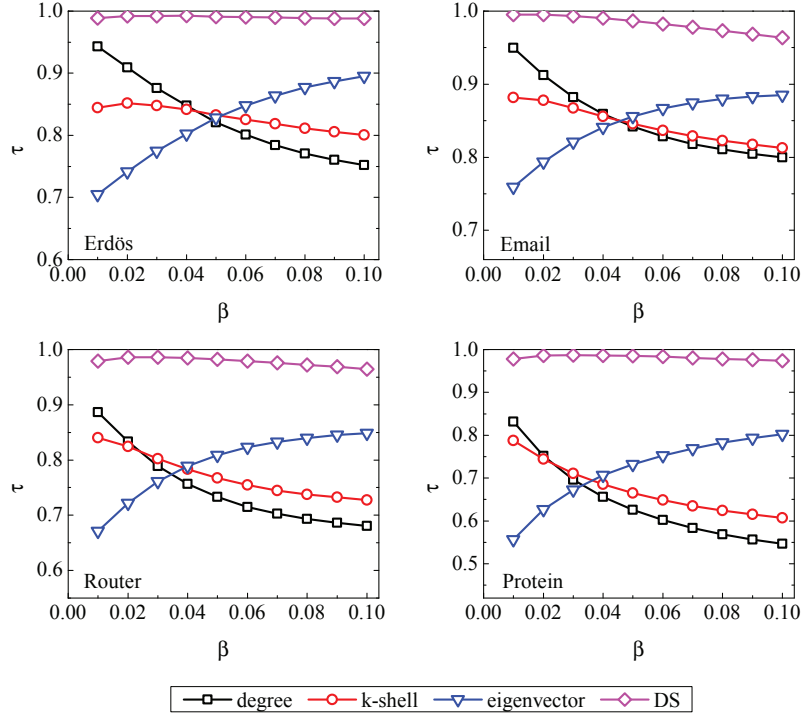}}
\caption{The accuracy of four centrality measures in evaluating nodes' spreading influences according to the SI model ($\mu=0$) in the four real networks, quantified by the Kendall's Tau. The spreading rate $\beta$ varies from 0.01 to 0.10, and the time step is set as $t=5$. Each data point is obtained by averaging over $10^4$ independent runs.}
\end{figure*}

\section{Discussion}

~\\

Estimating the spreading influences and then identifying the influential nodes are fundamental task before any regulation on the spreading process. For such task, most known works only took into account the topological information~\cite{Pei201302}. Recently, Aral and Walker~\cite{Aral2012} showed that the attributes of nodes are highly correlated with nodes' influences and tendencies to be influenced. In this paper, in addition to the topological information, we get down to the underlying spreading dynamics and propose a dynamics-sensitive (DS) centrality, which is a kind of weighted sum of walks ending at the target node, where both the spreading rate and spreading time are accounted in the weighting function. The DS centrality can be directly applied in quantifying the spreading influences of nodes, which remarkably outperforms degree, $k$-shell index and eigenvector centrality according to the empirical analyses of SIR model and SI model on four real networks. The DS centrality performs particular well in the early stage of spreading, which provides a powerful tool in early detection of potential super-spreaders for epidemic control.

The DS centrality tells us an often ignored fact that the most influential nodes are dependent not only on the network topology but also on the spreading dynamics. Given different models and parameters, the relative influences of nodes are also different. Roughly speaking, if the spreading rate is small, we can focus on the close neighborhood of a node since it is not easy to form a long spreading pathway (i.e., $\beta^t$ decays very fast as the increase of $t$ when $\beta$ is small) while if the spreading rate is high, the global topology should be considered. A clear limitation of this work is that before calculating the DS centrality, we have to know the spreading rate that is usually a hidden parameter. This parameter can be effectively estimated according to the early spreading process~\cite{Chen201401} and then we can calculate the DS centralities by varying the spreading rates over the estimated range and see which nodes are the most influential ones in average.

Some other centralities related to specific dynamical processes have also been proposed recently, including routing centrality~\cite{Dolev2010}, epidemic centrality~\cite{Sikic2013}, diffusion centrality~\cite{Ide2013}, percolation centrality~\cite{Piraveenan2013} and game centrality~\cite{Simko2013}. Comparing with these centralities, similar to the works by Klemm \emph{et al.}~\cite{Klemm201202,Ghanbarnejad2012,Klemm2013}, this paper provides a more general framework that could in principle deal with all networked Markov processes and thus can be extended and applied in many other important dynamics, such as the Ising model~\cite{Dorogovtsev2008}, Boolean dynamics~\cite{Kauffman1993}, voter model~\cite{Castellano2009}, synchronization~\cite{Arenas2008}, and so on. Furthermore, the DS centrality can also be directly extended to asymmetrical networks and weighted networks. We hope this work could highlight the significant role of underlying dynamics in quantifying the individual nodes' importance, and then the difference between lists of critical nodes may give us novel insights into the hardly notices distinguishing properties of different dynamical processes.

\section{Materials and Methods}

~\\

\subsection{Derivation of Eq. (1).}

The probabilities of nodes to be infected at time step $t=2$ can be approximated as
\begin{equation}
\textbf{x}(2)-\textbf{x}(1)=\beta\mathbf{A}\textbf{x}(1)+\beta\mathbf{A}(1-\mu)\textbf{x}(0)=\beta\mathbf{A}[\beta\mathbf{A}+(1-\mu)\mathbf{I}]\textbf{x}(0).
\end{equation}
We assume that when $t\leq p$, $\textbf{x}(p)-\textbf{x}(p-1)=\beta\mathbf{A}[\beta\mathbf{A}+(1-\mu)\mathbf{I}]^{p-1}\textbf{x}(0)$, then for $t=p+1$, we have
\begin{equation}
\begin{split}
\textbf{x}(p+1)-\textbf{x}(p)&=\beta\mathbf{A}\{\sum_{r=0}^{p-2}(1-\mu)^r[\textbf{x}(p-r)-\textbf{x}(p-r-1)]+(1-\mu)^{p-1}\textbf{x}(1)+(1-\mu)^p\textbf{x}(0)\}\\
&=\beta\mathbf{A}\{\sum_{r=0}^{p-2}(1-\mu)^r\beta\mathbf{A}[\beta\mathbf{A}+(1-\mu)\mathbf{I}]^{p-r-1}+(1-\mu)^{p-1}[\beta\mathbf{A}+(1-\mu)\textbf{I}]\}\textbf{x}(0)\\
&=\beta\mathbf{A}\{\sum_{r=0}^{p-3}(1-\mu)^r\beta\mathbf{A}[\beta\mathbf{A}+(1-\mu)\mathbf{I}]^{p-r-1}+(1-\mu)^{p-2}[\beta\mathbf{A}+(1-\mu)\textbf{I}]^2\}\textbf{x}(0)\\
&\cdots\\
&=\beta\mathbf{A}\{\beta\mathbf{A}[\beta\mathbf{A}+(1-\mu)\mathbf{I}]^{p-1}+(1-\mu)[\beta\mathbf{A}+(1-\mu)\textbf{I}]^{p-1}\}\textbf{x}(0)\\
&=\beta\mathbf{A}[\beta\mathbf{A}+(1-\mu)\mathbf{I}]^{p}\textbf{x}(0).\\
\end{split}
\end{equation}
Therefore, according to the mathematical induction, Eq. (1) is established. 

\subsection{Spreading Model.}

Here we apply the standard susceptible-infected-recovered (SIR) model
(also called the susceptible-infected-removed model) \cite{Hethcote2000}. In the SIR model, there are three kinds of individuals: (i) susceptible individuals that could be infected, (ii) infected individuals having been infected and being able to infect susceptible individuals, and (iii) recovered individuals that have been recovered and will never be infected again. In this paper, the spreading process starts with only one seed node being infected initially, and all other nodes are initially susceptible. At each time step,
each infected node makes contact with its neighbors and each susceptible
neighbor is infected with a probability $\beta$. Then each infected node
enters the recovered state with a probability $\mu$. Without loss of
generality, we set $\mu=1$. In the standard SI model, nodes can only be susceptible or infected, corresponding to the case with $\mu=0$.

\subsection{Benchmark Methods.}


The degree of an arbitrary node $i$ is defined as the number of its neighbors, namely
\begin{equation}\label{equation5}
k_{i}=\sum_{j=1}^{n}{a_{ij}},
\end{equation}
where $a_{ij}$ is the element of matrix $\mathbf{A}$. Degree is widely applied for its simplicity and low computational cost, which works especially well in evaluating nodes' spreading influences when the spreading rate is small.

The main idea of eigenvector centrality is that a node's importance is not only determined by itself, but also affected by its neighbors' importance~\cite{Borgatti2005}. Accordingly, eigenvector centrality of node $i$, $v_{i}$, is defined as
\begin{equation}\label{equation6}
v_{i}=\frac{1}{\lambda}\sum_{j=1}^{n}{a_{ij}v_{j}},
\end{equation}
where $\lambda$ is a constant. Obviously, Eq. (13) can be written in a compact form as
\begin{equation}\label{equation7}
\mathbf{A}\textbf{v}=\lambda\textbf{v},
\end{equation}
where $\textbf{v}=(v_1,v_2,\cdots,v_n)^T$. That is to say, $\textbf{v}$ is the eigenvector of the adjacent matrix $\mathbf{A}$ and $\lambda$ is the corresponding eigenvalue. Since the influences of nodes should be nonnegative, according to Perro-Frobenius Theorem \cite{Hom1985}, $\textbf{v}$ must be the largest eigenvector of $\textbf{A}$, say $\textbf{v}=\textbf{q}_{1}$.

Kitsak {\textit{et al.}} \cite{Kitsak201001} argued that $k$-shell index (i.e., coreness) is a better index than degree to locate the influential nodes. The $k$-shell can be obtained by the so-called $k$-core decomposition~\cite{Seidman1983}. The $k$-core decomposition process is initiated by removing all
nodes with degree $k=1$. This causes new nodes with degree $k\leq1$ to
appear. These are also removed and the process is continued until all remaining nodes are of degree $k>1$.  The removed nodes (together with associated links) form the 1-shell, and their $k$-shell indices are all one. We next repeat this pruning process for the nodes of degree $k=2$ to extract the 2-shell, that is, in
each step the nodes with degree $k\leq 2$ are removed. We continue with
the process until we have identified all higher-layer shells and all network
nodes have been removed. Then each node $i$ is assigned a $k$-shell index
$c_i$.

\subsection{Kendall's Tau.}

For each node $i$, we denote $y_i$ as its spreading influence and $z_i$ the target centrality measure (e.g., degree, $k$-shell index, eigenvector centrality and DS centrality), the accuracy of the target centrality in evaluating nodes' spreading influences can be quantified by the Kendall's Tau \cite{Kendall1938}, as
\begin{equation}\label{equation4}
\tau=\frac{2}{n(n-1)}\sum_{i<j}{\rm{sgn}}[(y_{i}-y_{j})(z_{i}-z_{j})],
\end{equation}
where sgn$(y)$ is a piecewise function, when $y>0$, sgn$(y)=+1$; $y<0$, sgn$(y)=-1$; when $y=0$, sgn$(y)=0$. $\tau$ measures the correlation between two ranking lists, whose value is in the range $[-1,1]$ and the larger $\tau$ corresponds to the better performance.

\subsection{Data description.}

Four real networks are studied in this paper as follows. (i) Erd\"{o}s, a scientific collaboration network, where nodes are scientists and edges represent the co-authorships. The data set can be freely downloaded from the web site http://wwwp.oakland.edu/enp/thedata/. (ii) Email~\cite{Guimera2003}, which is the email communication network of University Rovira i Virgili (URV) of Spain, involving faculty members, researchers, technicians, managers, administrators, and graduate students. (iii) Router~\cite{Rossi2015}, the Internet at the router level, where each node represents a router and an edge represents a connection between two routers. (iv) Protein \cite{Rual2005}, an initial version of a proteome-scale map of human binary protein-protein interaction. Basic statistical properties of the above four networks are presented in Table 1.

\begin{table}
\caption{Basic statistical features of Erd\"{o}s, Email, Router and Protein networks, including the number of nodes $n$, the number of the edges $e$, the average degree $\langle k\rangle$ and the reciprocal of the largest eigenvalue $1/\lambda_{1}$.}
\begin{center}
\begin{tabular} {l c c c c}
  \hline \hline
    Network         &$n$      &$e$       & $\langle k\rangle$       &$1/\lambda_{1}$       \\ \hline
   Erd\"{o}s       &456      &1314      &5.763                            &0.079        \\
   Email           &1133     &5451      &9.622                            &0.048         \\
   Router          &2114     &6632      &6.274                            &0.036      \\
   Protein         &2783     &6007      &4.317                            &0.063   \\
\hline \hline
\end{tabular}
\end{center}
\end{table}

\begin{addendum}

\item  The authors acknowledge the valuable discussion with Wen-Xu Wang and Liming Pan, as well as the Erd\"os Number Project in Oakland University for the data set. This work is supported by the National Natural Science Foundation of China (Nos. 71171136, 61374177, 112222543 and 61433014), the Shanghai Leading Academic Discipline Project of China (No. XTKX2012), JGL is supported by the Program for Professor of Special Appointment (Eastern Scholar) at Shanghai Institutions of Higher Learning.

\item[Author Contributions] JHL, JGL and TZ designed research, JHL, JGL and QG perform research, JHL, JGL and TZ analysed data, and JHL, JGL, QG and TZ wrote the paper. Correspondence and requests for materials should be addressed to J.G.L. (liujg004@ustc.edu.cn) or T.Z. (zhutou@ustc.edu).

\item[Competing Interests] The authors declare that they have no competing financial interests.

\end{addendum}
\ifthenelse{\boolean{SubmittedVersion}}{\processdelayedfloats}{%
\cleardoublepage}

\end{document}